\documentclass[twocolumn,aps,showpacs,amsmath,amssymb]{revtex4}
\usepackage{graphicx}
\usepackage[sort&compress]{natbib}

\begin{document}

\title{Mass distributions in a variational model}
\author{P. D. Stevenson and J. M. A. Broomfield}
\affiliation{Department of Physics, University of Surrey, Guildford,
  Surrey, GU2 7XH, UK}


\begin{abstract}
The time-dependent Hartree-Fock approach may be derived from a variational principle and a Slater Determinant wavefunction Ansatz.  It gives a good description of nuclear processes in which one-body collisions dominate and has been applied with success to giant resonances and collisions around the barrier.  It is inherently unable to give a good description of two-body observables.  A variational principle, due to Balian and V\'en\'eroni has been proposed which can be geared to good reproduction of two-body observables.  Keeping the Slater Determinant Ansatz, and restricting the two-body observables to be the squares of one-body observables, the procedure can be implemented as a modification of the time-dependent Hartree-Fock procedure.  Applications, using the Skyrme effective interaction, are presented for the mass distributions of fragments following de-excitation of the giant dipole resonance in $^{32}$S.  An illustration of the method's use in collisions is given.
\end{abstract}
\maketitle

\section{Introduction}

The nucleus is a complex many-body system which shows a rich variety of behaviour, including properties that can be identified as single-particle nature as well as collective motion.  Attempting to describe the nucleus in terms of its constituent proton and neutrons and their interaction is a difficult task, and one usually focuses on a restricted set of degrees of freedom that are of interest in the task at hand.

A venerable approximation that has a fairly general application is the self-consistent (or Hartree-Fock) mean-field approach, in which the particles are assumed to be moving freely (Pauli principle notwithstanding) in a potential generated by the average effect of the interaction of a given nucleon with the rest of the nucleons.  A vast body of literature exists on approaches starting from the mean-field point of view, using a variety of effective nuclear interactions.  For recent reviews see e.g. \cite{Sto07} for details of mean-field applications of the Skyrme interaction, \cite{Vre05} for relativistic models, and \cite{Ben03} for a combined review, including the Gogny force.  These reviews make clear the general applicability of these approaches, and how the mean field is a starting point for more elaborate theories that bring in correlations beyond the non-interacting single-particle picture.

Standard techniques for going beyond the mean-field approach make use of the Hartree-Fock solution as a reference state, from which combinations of configuration mixing and projection in various ways give rise to correlated states, and to spectra. Such methods include the Random Phase Approximation \cite{Ter07,Ber07}, the generator coordinate method \cite{Hee01,Klu08} and shell model-like configuration mixing approaches \cite{Pil06,Pil08}.  These methods generally rely on the outcome of static Hartree-Fock calculations to produce the variety of physical states.

To go beyond the mean-field picture in the time-dependent extension of the Hartree-Fock method, a different approach is usually used.  Hartree-Fock can be seen, as well as a lowest-order diagram calculation from a perturbation theory point of view, as a variational principle, and one may instead think about using an alternative variational principle to go beyond the limit of non-interacting particles.  The same is true of Time-Dependent Hartree-Fock, whose equations can be derived from a variation of the action.  In a series of papers, Balian and V\'en\'eroni developed a general variational principle which included as a special case the time-dependent Hartree-Fock approach, but which could be extended in various ways \cite{Bal85,Bal88,Bal92,Bal99}.  This form of beyond-mean-field approach leads to a natural way to work with time-dependent problems. 

Applications are made, in the present work, to calculations of number fluctuation following decay of giant resonance and reaction residues.  The method is reviewed in section \ref{sec:BV}, some issues of implementation discussed in section \ref{sec:numerical} and results in section \ref{sec:res}.  For further details, readers are referred to previous work \cite{Bro08,Bro09}.

\section{The Balian-V\'en\'eroni Method\label{sec:BV}}
The Time-dependent Hartree-Fock (TDHF) equations can be derived (amongst myriad other ways) from the variation of the time-dependent action

\begin{equation}
\delta\int_{t_1}^{t_2}\langle\Psi|i\partial_t-H|\Psi\rangle\,{\mathrm d}t =0
\end{equation}

whence the supposition of the state vector being of the form of a Slater Determinant containing a set of time-dependent single-particle states, $\{\phi\}$, yields the usual TDHF equations

\begin{eqnarray}
i\hbar\frac{\partial }{\partial t}\phi_j(r,t) &=& -\frac{\hbar}{2m}\nabla^2\phi_j(r,t)+\bar{v}(r,t)\phi_j(r,t)\nonumber \\
&-&\sum_k\phi_k(r,t)\int\mathrm{d}r'\phi^*_k(r',t)v(r-r')\phi_j(r',t).\nonumber\\
\label{eq:tdhf}
\end{eqnarray}

Choosing a Slater Determinant for the wavefunction, or equivalently a one-body density matrix satisfying $\rho^2=\rho$ results in an optimal treatment of the dynamics when one is concerned with one-body observables, with the choice of variational principal giving rise to a collective path which follows the minimum energy (within the constraints of the wavefunction Ansatz).  There is no reason to expect that any other two-body observables be well reproduced under these circumstances.  If one considers as a simple two-body observable the fluctuation in particle number

\begin{equation}
\langle\Delta N\rangle^2 = \langle N^2\rangle - \langle N \rangle^2
\end{equation}

one can demonstrate \cite{Das79} that TDHF can only achieve a theoretical upper limit of

\begin{equation}
\left.\Delta N_{\mathrm max}^2\right|_t = \left.\langle N\rangle\right|_t\left(1-\frac{1}{A}\left.\langle N\rangle\right|_t\right), \label{eq:dasso}
\end{equation}
which is a limit of the method, though the physical mass fluctuation may (and does) exceed this.  In their work \cite{Bal85} Balian and V\'en\'eroni tackled this issue, formulating instead a variational principle which sought to include the observable of interest in the variational space, so that one might trust calculations for at least that observable.  Their arguments lead to finding stationary values of a characteristic function

\begin{eqnarray}
J&=&\mathrm{Tr}\left[\hat{A}(t_1)D(t_1)\right]\nonumber\\&&-\int_{t_0}^{t_1}\mathrm{d}t\left( \mathrm{Tr} \left[\hat{A}(t)\frac{\mathrm{d}D(t)}{\mathrm{d}t}\right]-h(\hat{A}(t),D(t))\right)
\end{eqnarray}

in which $\hat{A}=\exp(-\epsilon \hat{N})$, where $\hat{N}$ is the one-body observable whose square is of interest, whose value at $t_1$ is required, $D$ is a density matrix, whose value at $t_0$ is known, $\epsilon$ is a small parameter and $h$ is a pseudo-Hamiltonian, defined as

\begin{equation}
h(\hat{A}(t),D(t))=-i\mathrm{Tr}\left(\hat{A}(t)\left[\hat{H}(t),D(t)\right]\right).
\end{equation}

In the special case that one restricts $D$ to be a one-body density matrix, one obtains finally an expression for the fluctuation of that operator

\begin{equation}
\Delta N^2(t_1) = \lim_{\epsilon\to0}\frac{1}{2\epsilon^2}\mathrm{Tr}\left[(\rho(t_0)-\sigma(t_0))^2\right] \label{eq:bvvar}
\end{equation}

where 

\begin{equation}
\sigma(t_1) = e^{i\epsilon\hat{N}}\rho(t_1)e^{-i\epsilon\hat{N}}. \label{eq:sigmabc}
\end{equation}

In these equations, $\rho$ and $\sigma$ are one-body densities which evolves according to the TDHF equations (\ref{eq:tdhf}). $\sigma$ is subject to the boundary conditions (\ref{eq:sigmabc}) at $t_1$.  In practice, one must perform a TDHF calculation forwards in time for the reaction of interest until time $t_1$, apply the transformation (\ref{eq:sigmabc}), then evolve $\sigma$ backwards in time to $t_0$ to evaluate (\ref{eq:bvvar}).

\section{Implementation\label{sec:numerical}}
The implementation of the Balian-V\'en\'eroni (BV) technique took as its starting point a three-dimensional time-dependent Hartree-Fock code which uses the full Skyrme interaction, including time-odd terms and the extended spin-orbit force.  It has recently been applied to giant resonances \cite{Mar05,Rei07} and collisions \cite{Mar06,Guo07}.  Since the first semi-realistic nuclear TDHF calculations \cite{Dav85}, advances in computing have recently allowed a more extensive and realistic series of calculations \cite{Mar06,Nak05,Sim03,Uma06}.  Since the BV technique is more computationally intensive than TDHF realistic calculations have previously been difficult, though previous calculations have been made of monopole resonances \cite{Tro85} and reactions \cite{Mar85}, though always with some restriction in either spatial symmetry or the effective interaction.

The Balian-V\'en\'eroni calculations drive the TDHF code by running from $t_0$ to $t_1$, applying the transformation (\ref{eq:sigmabc}) to the wavefunctions, then running back from $t_1$ to $t_0$.  This process is repeated for several values of the parameter $\epsilon$ in order to take the limit (\ref{eq:bvvar}).  The initial conditions at $t_0$ are set up for the collective mode of interest.   In the case of a giant dipole resonance, the starting wavefunctions are obtained from the results of a static Hartree-Fock calculation, modified by an instantaneous boost of the form

\begin{equation}
\phi(t_0) \rightarrow e^{iF_\tau f(r)\vec{k}\cdot\vec{r}} \phi(t_0)
\end{equation}

where $F_\tau = -1/N$ for neutron orbitals and $1/Z$ for proton orbitals. $f(r)$ is a spatial formfactor to limit the boost to the region of the nucleus.  

\begin{figure*}[tbh]
\centerline{\includegraphics[width=15cm]{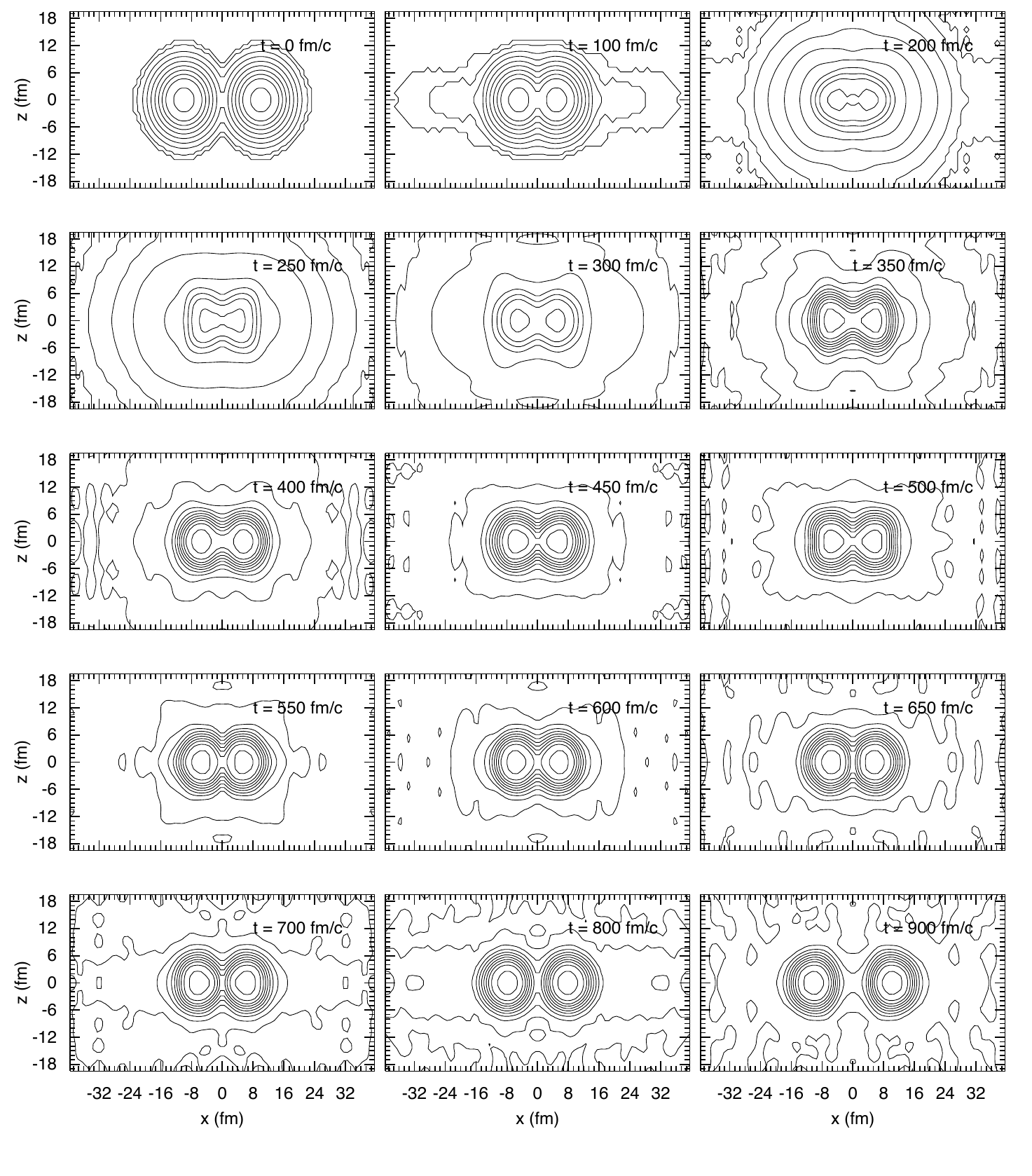}}
\caption{Logarithmic plot of the total particle density in a head-on collision of two $^{40}$Ca nuclei at a lab energy of 278 MeV.  Note that the starting position for the collision involves a slight overlap of the nuclei.  This is exaggerated by the logarithmic scale and the nucleon density at $(0,0)$ is of order $10^{-5}\rho_c$ where $\rho_c$ is the density at the centre of one of the nuclei.  The decay by emission of particles can be seen following the collision. \label{fig:ca40ca40headon}}
\end{figure*}

\begin{table}
\centerline{
\begin{tabular}{cccccc}
\hline
$R_c$ [fm] & $\langle N\rangle$ & $\Delta N_{\mathrm{TDHF}}$ & $\Delta (N_{\mathrm{TDHF}})_{\mathrm{max}}$ & $\Delta N_{\mathrm{BV}}$ & Change \\
\hline
8.0 & 26.637 & 2.022 & 2.113 & 2.351 & +16\% \\
8.5 & 26.734 & 2.009 & 2.098 & 2.330 & +16\% \\
9.0 & 26.898 & 1.988 & 2.071 & 2.292 & +15\% \\ \hline
\end{tabular}
}
\caption{Dependence of the bounding radius, $R_c$, separating the nucleus from the environment, on the number of nucleons in the residual nucleus, $\langle N\rangle$, the TDHF value of the particle number fluctuations, $\Delta N_{\mathrm{TDHF}}$, the theoretical maximum value that TDHF can give (\ref{eq:dasso}), $\Delta (N_{\mathrm{TDHF}})_{\mathrm{max}}$, the Balian-V\'en\'eroni fluctuation, $\Delta N_{\mathrm{BV}}$ and the change from TDHF to BV.\label{tab:rctest}}
\end{table}

Splitting up space into different regions - the nucleus and the environment - is key to the Balian-V\'en\'eroni process.  The wavefunctions are represented in a discretised box in space.  The box is divided into regions in a way specified by the problem at hand.  For a single nucleus in a giant resonance state, decaying by particle emission, a spherical region centered around the nuclear centre of mass is considered to be the nucleus, while the rest of the box is the environment.  This allows for both the number of emitted particles to be followed and defined, and for the transformation (\ref{eq:sigmabc}) to be applied only to the final nucleus and not the emitted flux.  Dependence of the results of the procedure on the size of the region holding the nucleus is one of the several checks for stability that has been made.

Table \ref{tab:rctest} shows the results of the variation of the cutoff radius for the case of a giant dipole resonance in $^{32}$S.  Clearly there is some dependence on the cutoff, as one would expect.  Increasing the size of this radius increases the number of final particles counted in the nucleus.  However, the changes are small compared with the number of dripped particles (i.e. around five) and the size of the fluctuations changes rather little.  The increase in the fluctuation in going from TDHF to BV is also extremely consistent.   A large amount of testing work has checked consistency as a function of geometry of cutoff, final running time $t_1$, method of extrapolation of the limit $\epsilon\to0$ and discretisation of the box.  For further details, see \cite{Broth}.

The implementation of collisions relies on begining with two static HF calculations to give the two colliding nuclei.  These are placed in the box, and a Galilean boost applied to both to set them on a collision path with the desired geometry.   Following collision, a single fragment is identified, the BV transformation applied, and time run backwards as usual.

As a final example of an issue of implementation, Figure \ref{fig:ca40ca40headon} shows a logarithmic plot of the nuclear density during the head-on collision of two $^{40}$Ca nuclei.  The plot shows that following the collision, the excitation of the compound nucleus results in the emission of particles, seen as the outgoing waves in the third and fourth frames.  One also sees that at later times the nuclei are surrounded by a gas of nucleons formed by reflection of the outgoing flux from walls of the box.  This unphysical reflection is in principle a source of error in the BV technique since the interference of the reflected flux with the nuclear region can give perturbations to the observables.  Techniques are available which can eliminate the flux \cite{Rei06,Man98}, but they are either computationally costly or are not irreversible, or both.

\section{Results\label{sec:res}}

\begin{figure}
\centerline{\includegraphics[width=9cm]{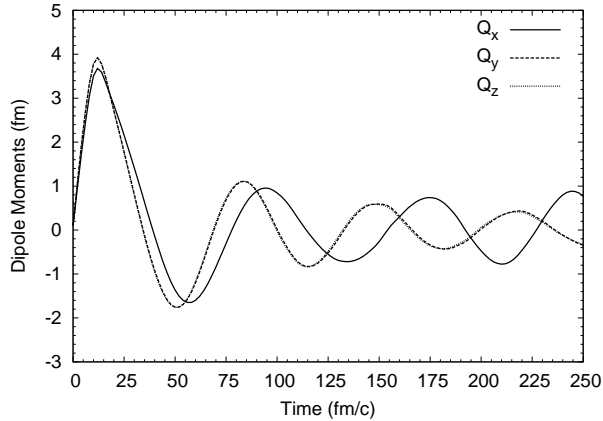}}
\caption{Time-dependent dipole moments of $^{32}$S following a dipole boost.\label{fig:s32}}
\end{figure}

Following the prescription of the previous chapter, some results of giant dipole resonances in $^{32}$S and collisions between two $^{16}$O nuclei are presented.  Figure \ref{fig:s32} shows the dipole repsonse in $^{32}$S, as calculated with the SLy6 Skyrme parameterisation \cite{Cha97} following an isovector dipole boost.  The moments in the $y$- and $z$- directions are superimposed since the nucleus is axially symmetric with a prolate deformation.  The $x$-axis is the major axis and hence the giant resonance has a slower mode in the $x$ direction, and two faster modes in the $y$ and $z$ directions, as expected from a prolate nucleus.    

Following the deexcitation, the BV transformation is applied within the cutoff radius, and the modified density matrix evolved back to $t_0$.  Results are presented in Table \ref{tab:skyrmes} for the fluctuations in particle number for this calculation, as well as the equivalent calculation with different Skyrme parameterisations.

\begin{table}[th]
\begin{tabular}{lcccc}
\hline
Skyrme force & $\langle N\rangle$ & $\Delta N_{\mathrm{TDHF}}$ & $\Delta N_{\mathrm{BV}}$ & Change \\
\hline
SkM* & 26.172 & 2.101 & 2.287 & +9\% \\
SLy4 & 26.293 & 2.060 & 2.537 & +23\% \\
SLy4d& 26.421 & 2.053 & 2.463 & +20\% \\
SLy6 & 26.637 & 2.022 & 2.351 & +16\% \\
\hline
\end{tabular}
\caption{Number fluctuation following dipole excition in $^{32}$S using Skyrme force parameterisations SkM* \cite{skms}, SLy4 \cite{Cha97}, SLy4d \cite{sly4d} and SLy6 \cite{Cha97}.\label{tab:skyrmes}}
\end{table}

The variation in predictions across different Skyrme forces exceeds that due to any of the numerical uncertainties discussed briefly in the previous section.  This is reassuring since it means that this method is sensitive to properties of the effective interaction, and the predictions of fluctuations may be able to be used in future to further constrain and select between different Skyrme forces, adding to the increasing range of properties which are used to judge Skyrme parameterisations \cite{Rik03}.  It should be pointed out that Skyrme parameterisations are fitted essentially to the properties of the ground states of doubly-magic nuclei and calculations of, for example, giant resonances in open-shell nuclei are essentially parameter-free.

As shown in Figure \ref{fig:ca40ca40headon}, calculations of collisions have also been performed. In this present paper, no analysis is presented beyond the discussion of the figure in the previous section. Work on completing the analysis is in progress, and will be reported elsewhere.  The first analysis of collisions between $^{16}$O nuclei can be found in the literature \cite{Bro09}.

\section{Conclusion\label{sec:conc}}

The Balian-V\'en\'eroni method for reliably evaluating the fluctuations of one-body observables has been implemented for symmetry-unrestricted nuclear dynamics with the Skyrme interaction.  First results have been presented for giant resonances and collisions.  The approach is computationally intensive and has some inherent numerical issues involved in its implementation, but these are well in control, and the differences evident in different Skyrme parameterisations are larger than the numerical uncertainty.  The Balian-V\'en\'eroni technique can therefore be used both to calculate fluctuations, as well as to serve as a filter for Skyrme interactions.

\acknowledgments
The authors acknowledge useful discussions with P.-G. Reinhard and J. A. Maruhn concerning the TDHF code, and Ph. Chomaz for suggesting re-visiting the Balian-V\'en\'eroni method.  Support is gratefully acknowledged from the UK Science and Technology Facilities Council (STFC).

\end{document}